\documentstyle[prb,aps,twocolumn,epsf]{revtex} 
\begin{document}
\title{Magnetic anomalies in the spin chain system,
Sr$_3$Cu$_{1-x}$Zn$_x$IrO$_6$}
\author{Asad Niazi, E.V. Sampathkumaran,$^a$ P.L. Paulose, D. Eckert,$^*$
A. Handstein$^*$, and K.-H. M\"uller$^*$}
\address{Tata Institute of Fundamental Research, Homi Bhabha Road, Colaba,
Mumbai - 400005, INDIA\\
$^*$Institut f\"ur Festk\"orper- and Werkstofforschung Dresden, Postfach
270016, D-01171 Dresden, GERMANY\\
$^a$E-mail address: sampath@tifr.res.in}
\maketitle
\begin{abstract}

We report the results of ac and dc magnetization (M) and
heat-capacity (C) measurements on the solid solution,
Sr$_3$Cu$_{1-x}$Zn$_x$IrO$_6$. While the Zn end member is known to
form in a rhombohedral pseudo one-dimensional K$_4$CdCl$_6$ structure with
an antiferromagnetic ordering temperature of (T$_N$ =) 19 K, the Cu end
member has been reported to form in a monoclinically distorted form with a
Curie temperature of (T$_C$ =) 19 K.  The magnetism of the Zn compound is 
found to be robust to synthetic conditions and is broadly
consistent with the behavior known in the
literature. However, we find a lower magnetic ordering
temperature (T$_o$) for our Cu compound ($\sim$ 13 K), thereby suggesting that T$_o$ is
sensitive to synthetic conditions. The Cu sample appears to be in a spin-glass-like
state at low temperatures, judged by a frequency dependence of ac magnetic
susceptibility and a broadening of the C anomaly at the onset of magnetic
ordering, in sharp contrast to earlier proposals. Small applications of magnetic field,
 however, drive this system to
ferromagnetism as inferred from the M data.  Small substitutions for Cu/Zn 
(x= 0.75 or 0.25) significantly depress
magnetic ordering; in other words, T$_o$ varies non-monotonically with x
(T$_o$ $\sim$ 6, 3 and 4 K for x= 0.25, 0.5, and 0.67
respectively). The plot of inverse susceptibility versus temperature is
non-linear in the paramagnetic state as if correlations within (or among)
the magnetic chains continuously vary with temperature.  The results
establish that this class of oxides is quite rich in physics.

\end{abstract}

\vskip 0.5cm
{PACS Nos. 75.50.-y; 75.20. Ck; 75.30 Cr}
%\newpage

\maketitle The investigation of compounds with low-dimensional
crystallographic features has gained considerable momentum in recent
years, following a variety of physical phenomena such systems are found to
exhibit. For instance, the magnetic systems of low dimensionality with
integer spins and non-integer spins are expected to behave differently in
the sense that the former exhibits a (Haldane) gap, while latter type does
not.\cite{1} It is therefore worthwhile to identify new systems with low
dimensional structural features and to subject them to intense
investigations. 

In this regard, the compounds of the type, A$_3$MXO$_6$
(A= Ca, Sr, Ba; M= a magnetic or a non-magnetic ion; X= a magnetic ion)
are of considerable interest (see, for instance, Refs 2-20), as this class
of compounds crystallize in a rhombohedral structure of K$_4$CdCl$_6$ type
(space group R$\bar3$c) characterized by the presence of psuedo-one
dimensional chains of M-X along c-direction; the structure consists of
infinite chains of alternating face-sharing MO$_6$ trigonal prisms and
XO$_6$ octahedra; the one-dimensional character can be expected under
favorable conditions (that is, if intra-chain magnetic ion separation is
shorter than that of interchain) as these chains are separated by
non-magnetic A ions. The ions at the M
and X sites are generally different, except the case of Co which occupies
both the sites simultaneously, e.g., Ca$_3$Co$_2$O$_6$ 
(Ref. 11). 
 Both M and X can be magnetic ions or M can be a non-magnetic ion 
(like Li, Na, K), while X can be a magnetic-moment
carrying ion. The variety of substitutions at these two sites (M and X)
this structure offers provide interesting opportunities to tune interchain
and intrachain interactions by selective substitution by a magnetic or a
non-magnetic ion at the M site, resulting in different types of magnetic
structure. 

We believe that this class of compounds have not been
sufficiently studied in spite of the fact that the reports available 
to date reveal interesting anomalies: For instance, the series
Sr$_3$CuPt$_{1-x}$Ir$_{x}$O$_6$, in which initially 
random quantum spin chain paramagnetism was  
proposed for x= 0.5, has been found to exhibit
more complex magnetism.\cite{15,16} Zn in Sr$_3$ZnIrO$_6$
exists in an unusual trigonal prismatic coordination.\cite{13} Co present
at the two sites (with high and low spin) possesses different magnetic
moments exhibiting multiple sharp transitions in the isothermal
magnetization (M) data.\cite{11} Magnetic ordering is seen for 4d and 
even for 5d ions (like Pt, Ir, Rh), and particularly noteworthy is that Ru (at the
X site) exhibits magnetic ordering at an unusually large temperature
(about 110 K) with both ferromagnetic and antiferromagnetic-like features
in the M data.\cite{3,4,18,19} In addition, due to Jahn-Teller effect of
the Cu ions (at the M site), for Cu containing compounds, the
one-dimensional structure is not that of a simple collinear
chain. For instance, in Sr$_3$CuIrO$_6$, the Ir ions lie on a
straight line, whereas the Cu$^{2+}$ ions are offset from the chain axis
toward one of the faces of the oxygen trigonal prism by 0.53 \AA\ resulting
in a pseudosquare planar coordination; the direction of offset rotates by
180$^o$ from one Cu atom to the next along the chain. These displacements
of atoms result in a monoclinic distortion (space group C2/c) of the
structure.\cite{21} It is thus clear that these oxides exhibit a variety of
interesting magnetic and crystallographic features and it is therefore of
interest to carry out a systematic study of these compounds. We have
earlier reported interesting magnetic anomalies in Ru based
compounds\cite{18,19} and, as a continuation of our studies in this
direction, we have carried out magnetic investigations on the oxides,
Sr$_3$Cu$_{1-x}$Zn$_{x}$IrO$_6$.

	The investigation of the present Cu-Zn based solid solution was
primarily motivated by interesting magnetic anomalies in the solid
solution Sr$_3$CuPt$_{1-x}$Ir$_x$O$_6$ (Ref. 15,16) due to coexisting
ferro and antiferro-magnetic interactions within the chain. There are
considerable theoretical interests on the thermodynamics of systems with
random ferromagnetic-antiferromagnetic chains\cite{22} in the current
literature and therefore it is worthwhile to identify such systems.  The
Cu and Zn end members in the present solid solution have been previously
reported to order essentially ferro and antiferromagnetically respectively
close to 19 K.\cite{13,14,15,16} Therefore, this solid solution is
expected to provide an opportunity to probe the magnetic behavior by
disruption of the magnetic chains by substitution of Zn for Cu. Apart from
throwing light on the magnetism of end members, we observe many interesing
anomalies in our measurements which are reported in this article.

	The samples, Sr$_3$Cu$_{1-x}$Zn$_x$IrO$_6$, (x= 0.0, 0.25, 0.5,
0.67, 0.75 and 1.0), were prepared by a conventional solid state reaction
method. The stoichiometric amounts of high purity ($>$ 99.9\%) SrCO$_3$,
Ir metal, CuO and ZnO were intimately mixed under acetone and preheated at
800 $^\circ$C for 33 hours. These were thoroughly ground
again, pelletized and sintered at 1150
$^\circ$C for 100 hours with three intermediate grindings.  The samples
were characterized to be single phases by x-ray diffraction (Cu
K$_{\alpha}$). Rietveld refinement of the x-ray data using DBWS-9411 
(Ref. 23)  yielded the monoclinic structure for x= 0.0 (as reported earlier)
and 0.25. This is also evident  from the splitting of some of the lines (see, for instance, the
lines falling in range 42 to 46 degrees in Fig. 1). For x$\geq$ 0.5, the monoclinic 
distortion disappears and the patterns could be satisfactorily fitted within 
the rhombohedral structure. The lattice constants derived from the refinement
are included in Fig. 1 and the values for the end members agree with the literature.

We have measured dc
magnetic susceptibilty ($\chi$) as a function of temperature (T= 2--300 K)
at different magnetic fields (H) and isothermal magnetization (M) at 2 K,
by a commercial superconducting quantum interference device (Quantum
Design) as well as a vibrating sample magnetometer (Oxford
Instruments). In addition, ac $\chi$ measurements were performed in the
vicinity of the magnetic ordering temperature (T$_o$) at various frequencies.  
Heat-capacity (C) measurements (2--40 K) were performed by a semi-adiabatic
heat-pulse method.

	We first discuss the observations on the end members. With respect
to Sr$_3$CuIrO$_6$, we display low T dc $\chi$ in figure 2. It is
distinctly clear that there is a sharp rise of $\chi$ below 13 K for the
data recorded at different fields indicating the onset of magnetic
ordering. However, the observed T$_o$
is significantly lower than that reported (19 K) in the literature.\cite{15,16}
In order to resolve this issue, we prepared another batch of sample (batch
2) with the same heat treatment conditions, however with a reduced
duration of hours of sintering (1150 $^\circ$C, 72 hours with
three intermittent grindings); and we noticed\cite{24} that the value of
T$_o$ now agrees with that of literature. This clearly establishes that
T$_o$ is sample dependent for this composition and prolonging the
heat-treatment well beyond 24 hours (after each grinding) depresses T$_o$
- a fact not known earlier in the literature. We have confirmed this
observation by repeated preparation of samples.  A slight reduction
(5$\%$) in Cu in the starting stoichiometry also modifies T$_o$ in the
same way.\cite{24} In order to further understand the origin
of this sample dependence, we have carried out energy dispersive x-ray
(EDX) analysis. We notice that the atomic ratios of metallic components are very
close to the stoichiometry, but O contents are different in these two
specimens; batch 1 (the batch discussed in this article) contains excess
oxygen (close to atomic ratio 7), while in batch 2, O content is close to
the ratio 6. It is therefore clear that prolonged heat treatment results
in excess oxygen intake, resulting in a depression of T$_o$. It is of
interest to focus future studies on the site of extra oxygen. 

With respect
to the nature of the magnetic ordering, a plot of inverse $\chi$ versus T
at low temperatures (below 30 K, Fig. 3) extrapolates to a positive value
of paramagnetic Curie temperature ($\Theta_p$) as if the compound
undergoes ferromagnetic ordering. However, a look at the frequency
dependence of ac $\chi$ (Fig. 4), is quite revealing. Not only the peak
height, but also the peak temperature in the ac $\chi$ data is found to be
highly sensitive to frequency, shifting it upwards by as much as 1.6 K as
the ac frequency is increased from 3 to 1000 Hz. This property implies
that this compound behaves like a spin-glass, in sharp contrast to earlier
proposals in the literature. This frequency dependence is relatively less
(about 0.5 K) for the batch 2 specimen.\cite{24} Interestingly, this frequency
dependence vanishes by the small application of an external dc magnetic
field (say, 1 kOe). We interpret, therefore, that the spin-glass-like state
turns to ferromagnetic ordering by the presence of a small magnetic field.
The dc magnetization behavior  at 2 K, in fact, establishes 
field-induced ferromagnetism of
this compound (Figs. 5 and 6). 
A careful look at the hysteresis loops (Fig. 5) at 2 K suggests
that the initial virgin curve is not typical of perfect ferromagnets as 
there is a flat region for H$<<$500 Oe before M rises
sharply, followed by saturation at higher fields. This finding endorses
that there is actually an antiferromagnetic component in zero-field. 
Though domain wall pinning effects
also can  give rise to initial flattenning of M vs H plot, this possibility
is ruled out considering that the virgin curve in the hysteresis plot
deviates from the center of the hysteresis loop.  Another 
noteworthy feature (Fig. 2) in the magnetically ordered
state is that, there is a peak at about 9 K in the plot of
zero-field-cooled (ZFC) dc $\chi$ versus T for H= 10 Oe  decreasing to
 relatively
insignificant values of $\chi$ at about 2 K. This peak shifts to a lower 
temperature (5 K) for H= 100 Oe. Higher applications
of H (say, 5 kOe) wipe out the drop in $\chi$ at low temperatures.
 The field-cooled (FC)
curves, however, do not show this anomaly for any value of H.
Clearly, there is an interesting magnetic anomaly well below T$_o$ which is
history/field dependent.  Such low temperature anomalies were reported
even in the ac $\chi$ data earlier.\cite{16} It is worthwhile to focus
future studies on this feature.

Other important observations for the Cu sample are: 
(i) The isothermal M, subsequent to a
steep rise around 300 Oe, undergoes a very weak rise at higher fields (Fig.
6). The saturation moment value, derived from the
linear extrapolation of the high field M data to zero field  
is much smaller (close to 0.35 $\mu_B$) than that expected 
if one assumes ferromagnetically-coupled spin
1/2 at both Cu and Ir sites. This value is also smaller than that of the effective 
moment ($\mu_{eff}$= 2.5 $\mu_B$ per
formula unit) obtained from the paramagnetic high temperature (200--300 K) 
linear region of the inverse 
$\chi$ vs T plot. This observation, agreeing with that of Ref. 16, suggests that
interchain interaction could be antiferromagnetic.  (ii) The plot of
inverse $\chi$ versus T in the entire range of the paramagnetic state
(Fig. 3) is found to be highly non-linear. Since these compounds are
insulators, such a non-linearity cannot arise from  (temperature
independent) Pauli paramagnetic contribution. This interesting finding
therefore is attributed to variation of the  effective coupling with changing T.
This implies the formation of uncoupled magnetic segments\cite{24} along
the chain, probably isolated by disorder/vacancies at rather high
temperatures in the paramagnetic state. We also note that the degree of
the curvature decreases (Fig. 3, see  curve 0.0 b) if one takes excess CuO (say, 10\%)
in the starting stoichiometry, thereby establishing the role of
vacancies in the stoichiometric material on the $\chi$ behavior discussed
above. (iii) The FC and ZFC $\chi$ curves begin to deviate just
above the peak temperature, a characteristic feature in spin-glasses. 
(iv) The anomaly normally expected due to magnetic ordering at T$_o$ in
the plot of C versus T is absent (Fig. 7), which is a characteristic of
spin-glasses as well.
	
	In sharp contrast to the behavior of the Cu sample, the Zn sample
is found to be more robust both to slight deviations in stoichiometry as
well as to heat treatment conditions. Both the batches (1 and 2) of the stoichiometric samples
prepared under the same conditions as the Cu samples 
exhibit similar magnetic behavior: There exists a well-defined peak in
$\chi$ at about 20 K (Fig. 2), similar to that reported earlier.\cite{16}
There is a weak rise with decreasing T at low temperatures in our data 
which could arise
from paramagnetism of the moment-carrying Ir ions at the free ends of the
segments. We observe neither frequency dependence of ac $\chi$ nor
significant differences in ZFC-FC dc $\chi$ curves, unlike in the Cu sample,
 establishing that this compound is not a spin-glass. In fact, the
isothermal M at 2 K undergoes linear variation with H till the highest
field measured (120 kOe), indicating that the
Zn end member undergoes long range antiferromagnetic ordering, clearly
from Ir ions.  There is some degree of non-linearity in the plot of
inverse $\chi$ versus T in the paramagnetic state (Fig. 3), the curvature
of which gets reduced if one takes 10\% excess ZnO in starting
composition. This endorses our view stated above that vacancies along the
chain cause this effect. The $\Theta_p$ (in the range 30--50 K) is 
found to be negative with the
same magnitude as that of T$_o$ and thus both the sign and the magnitude
are in full agreement with the proposal of antiferromagnetic ordering in
this compound.  The value of $\mu_{eff}$ obtained from the linear region
(1.6 $\mu_B$) agrees with that reported earlier.\cite{16} It should also
be noted that there is a very prominent anomaly in C in the vicinity of
magnetic ordering (Fig. 7), unlike in the Cu sample, which establishes
long range magnetic ordering in this compound.

	It is quite instructive to see how the magnetism evolves as Cu is
gradually substituted for Zn. For this purpose, we present here the
results of our investigations on the compounds, x= 0.25, 0.5, 0.67 and
0.75 (batch 1) and the findings are found to be qualitatively similar even
for those prepared in batch 2. For x= 0.75, it is interesting to see that
the $\chi$ (both dc and ac) undergoes a monotonic increase with decreasing
temperature, as though there is no magnetic ordering down to 2 K. There is
no frequency dependence of ac $\chi$ as well in the T range of
investigation. However, in the dc $\chi$ data, there is a marginal change
in slope around 7 K (marked by a vertical arrow in Fig. 2), a feature
which is confirmed for repeated preparation of samples. In order to probe
the origin of this feature, we have measured C and the data (Fig. 6) shows
a weak, but noticeable drop around 7 K. This indicates  
an onset of weak magnetic order below 7 K. Considering the trends
in ac $\chi$ data discussed above, the magnetic ordering is not even of a
spin-glass type, which is also endorsed by the absence of a bifurcation of
ZFC-FC $\chi$ curves. It is possible that  this signals incipient
antiferromagnetism for this composition, considering that the next higher
Zn composition is antiferromagnetic.  In any case, it is clear that T$_o$
is depressed significantly by small substitutions of Cu/Zn, which could be
due to local magnetic frustration effects due to substitution of Cu for
Zn. In fact, a careful look at the plot of M versus H (Fig. 6) suggests
that the linearity observed for Zn end member is absent for this
composition and the plot exhibits a curvature as though there is a
ferromagnetic component. However, the plot is still non-hysteretic. It is
also worth noting that the inverse $\chi$ (Fig. 3) is non-linear - a feature
incidentally maintained for all the solid solutions.  Clearly, a
small replacement of Cu for Zn brings out dramatic changes in the magnetic
characteristics. 

Upon further replacement of Zn by Cu, say x= 0.67,
we observe features due to magnetic ordering in the $\chi$ data,
however only at a low temperature. The ZFC $\chi$ curve
exhibits a broad peak around 3 K with the ZFC-FC curves bifurcating below 4 K.
The ac $\chi$ data also show a tendency to flatten below 4 K without any
noticeable frequency dependence. The behavior of isothermal M at 2 K is
qualitatively similar to that of x= 0.75. On the basis of M behavior, one
can infer that this compound undergoes magnetic ordering of a
canted-moment type. 

For x= 0.5, there is a very distinct peak at 3 K in ac
$\chi$ data, undergoing an upward shift with increasing frequency - a
feature vanishing for small applications of H, say 1 kOe, as in the case
of the Cu end member. There are corresponding anomalies in the plot of dc
$\chi$ versus T plot as well: The ZFC-FC curves deviate from each other below
T$_o$.  We infer that Cu-end-member-like behavior sets in at
this composition, though T$_o$ is reduced. In contrast to the behavior of
the compositions with lesser Cu content discussed above, M at 2 K rises
rather sharply at low fields with a curvature towards saturation, 
though it continues to rise till the high fields measured (Fig. 6).
This observation also suggests the existence of a strong competition from
ferromagnetic correlations. It may be better to classify this composition
as a magnetically frustrated system, considering the absence of a
well-defined prominent peak in the C data around 3 K; a weak drop below 3
K is not characteristic of magnetic systems exhibiting long range order.

For x= 0.25, the features in ac and dc $\chi$, isothermal M, hysteresis
and the C data are similar to that of x= 0.0, except that the magnetic
ordering appears below 6 K, and this compound appears to be spin-glass-like
in zero-field; this value of T$_o$ also establishes that a small
disruption of the chain depresses magnetism at the Cu end as well.

Summarizing, we have perfomed magnetic measurements on an interesting
class of oxides  exhibiting quasi-one-dimensional structural
characteristics. 
We emphasize on the following findings: (i) The Cu end member exhibits
spin-glass-like characteristics in the ac $\chi$ data in zero magnetic
field, which reveals that the exchange interaction is actually
three-dimensional and not one-dimensional at least in the magnetically
ordered state, since in one-dimension it is rather difficult to visualise
spin-glass freezing.  (ii) The spin-glass-like features are observed for
all x $<$ 0.5 and an application of a magnetic field drives this state
towards a ferromagnetic state with a reduced moment.\cite{25} (iii) For
the Cu end, the T$_o$ is found to be sensitive to the preparative
conditions of the specimens presumably due to the deviations in oxygen/Cu
stoichiometry. However, the Zn end member does not exhibit such a
sensitivity of magnetism on the preparative conditions and thus the
antiferromagnetism observed in this compound is rather robust.  (iv) The T$_o$
is found to undergo a sharp dip for small substitutions of Cu/Zn and thus exhibits 
non-monotonic variation with x, implying strong competition between ferromagnetic
and antiferromagnetic interactions in the solid solution.  (v) The absence of Curie-Weiss
behavior of $\chi$ in the paramagnetic state  in the
entire solid solution indicates the existence of 
paramagnetic segments which are isolated from each other due to the disruption of chains by
disorder, with drastic changes in the effective magnetic couplings among these
segments with varying temperature.  The results reveal the existence of
an interesting T-H-x magnetic phase diagram for Sr$_3$Cu$_{1-x}$Zn$_x$IrO$_6$,
establishing richness in the physics\cite{26,27} of these compounds.

 Acknowledgements:  A part of this work, performed in Dresden, was supported
 by the Deutsche Forschungsgemeinschaft
 within the SFB463. One of us (EVS) would like to thank Prof. C. Laubschat
 for an invitation to Dresden. We thank  Dr. Sudhakar Reddy for EDX analysis.
The help of Kartik K. Iyer while preparing
samples is acknowledged.

%\newpage

\begin{figure}
\centerline{\epsfxsize=7cm\epsfbox{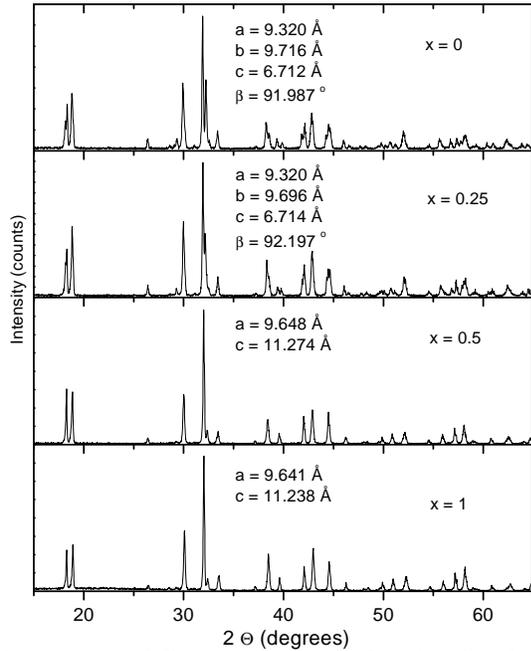}}
\caption{X-ray diffraction pattern (Cu K$_\alpha$) for the compounds of
the series, Sr$_3$Cu$_{1-x}$Zn$_x$IrO$_6$. For x$>$0.5, the patterns are
similar and hence not shown for all compositions.  For x = 0.0 and 0.25,
monoclinic distortion leads to the splitting of the lines, which is clearly resolved for 
the most intense line and the ones in the range 42 to 46 degrees.}
\end{figure}

\begin{figure}
\centerline{\epsfxsize=8cm\epsfbox{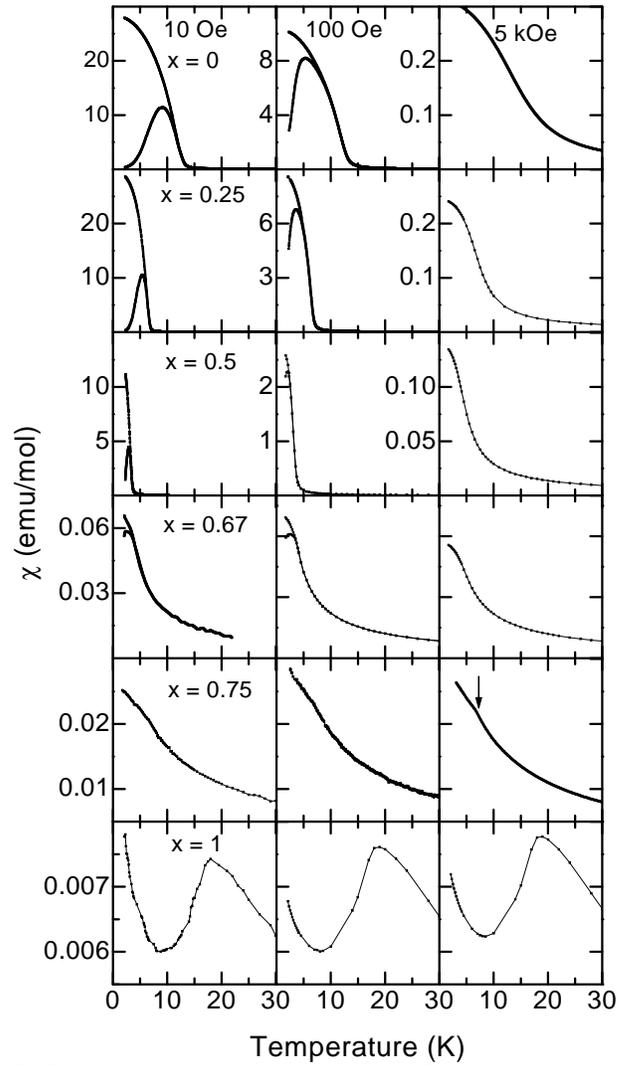}}
\caption{Temperature dependence of dc magnetic susceptibility ($\chi$) below
30 K measured in the presence of different magnetic fields (10, 100 and
5000 Oe) for the zero-field-cooled (ZFC) and field-coled (FC) states of
the specimens, Sr$_3$Cu$_{1-x}$Zn$_x$IrO$_6$. The vertical arrow marks the
temperature at which there is a sudden change in slope for x =  0.75. The
plots in a given row are pertinent to one composition.}

\end{figure}
\newpage

\begin{figure}
\centerline{\epsfxsize=6cm\epsfbox{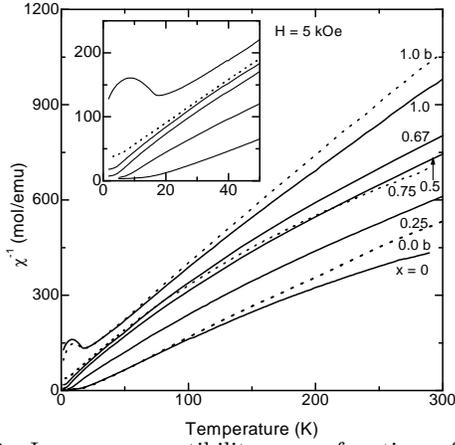}}
\caption{Inverse susceptibility as a function of temperature measured in
the presence of a magnetic field of 5 kOe for all the compositions of the
series, Sr$_3$Cu$_{1-x}$Zn$_x$IrO$_6$ (zero-field-cooled state).  The
data labelled, 0.0 b and 1.0 b, shown by dotted lines corresponding to Cu
and Zn end members with 10\% excess CuO and ZnO respectively in the
stoichiometry, are also shown to highlight reduced curvature of the plots
in the paramagnetic state for these compositions. The insets show the data 
for T $<$50 K for all the compositions (except those with 
excess CuO/ZnO) to highlight the sign and the magnitude of the 
Curie-Weiss parameter at low temperatures.}
             
\end{figure}

\begin{figure}
\centerline{\epsfxsize=7cm\epsfbox{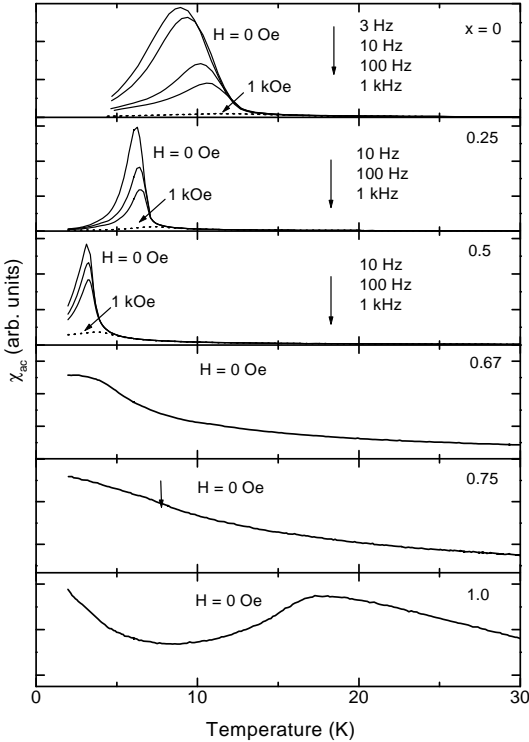}}
\caption{Real part of ac susceptibility as a function of temperature for
the series, Sr$_3$Cu$_{1-x}$Zn$_x$IrO$_6$ at different frequencies. The
plots at different frequencies for x = 0.67, 0.75 and 1.0 overlap. For x =
0.0, 0.25 and 0.5, the curves are shown also in the presence of a magnetic
field of 1 kOe and these at different frequencies overlap with each other.}

\end{figure}

\begin{figure}
\centerline{\epsfxsize=6cm\epsfbox{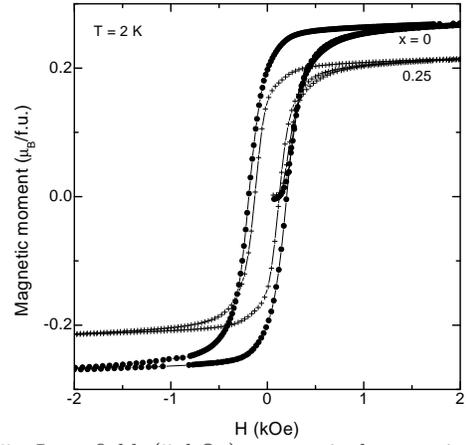}}
\caption{Low field (5 kOe) magnetic hysteresis loops at 2 K for two
compositions x = 0.0 and 0.25 in the series, Sr$_3$Cu$_{1-x}$Zn$_x$IrO$_6$.
Since no hysteresis has been observed for other compositions, the
corresponding curves are not being shown here.}

\end{figure}
\vskip3cm
\begin{figure}
\centerline{\epsfxsize=7cm\epsfbox{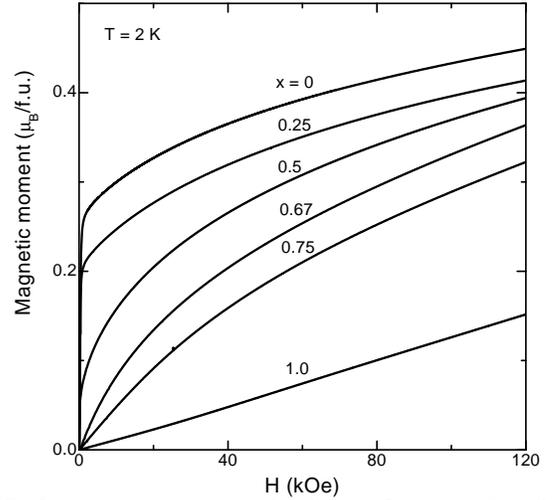}}
\caption{Isothermal magnetization at 2 K as a function of magnetic field
up to 120 kOe for all the samples of the series,
Sr$_3$Cu$_{1-x}$Zn$_x$IrO$_6$.}

\end{figure}
\newpage
\begin{figure}
\centerline{\epsfxsize=7cm\epsfbox{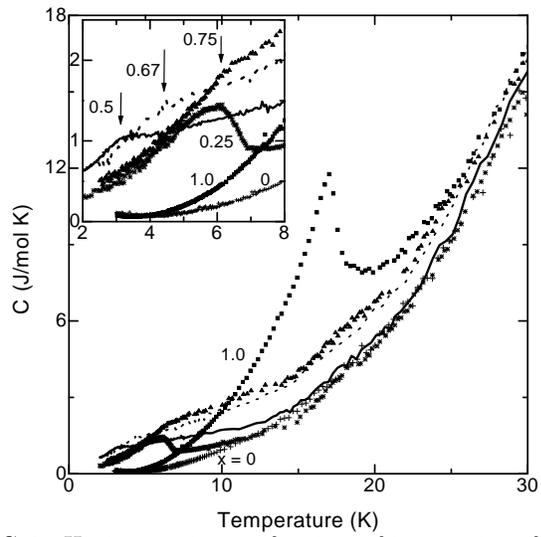}}
\caption{Heat capacity as a function of temperature for the series, 
Sr$_3$Cu$_{1-x}$Zn$_x$IrO$_6$. The inset shows
the data at low temperatures in an expanded form so as to highlight the
weak features (marked by vertical arrows) in some cases due to magnetic
ordering.}

\end{figure}      

\end{document}